# AI Biases as Asymmetries: A Review to Guide Practice

Gabriella Waters (CEAMLS, Morgan State University)*
Phillip Honenberger (CEAMLS, Morgan State University)*

*Equal contribution
[Preprint – Nov. 21, 2024]

**Abstract**

The understanding of bias in AI is currently undergoing a revolution. Initially understood as errors or flaws, biases are increasingly recognized as integral to AI systems and sometimes preferable to less biased alternatives. In this paper we review the reasons for this changed understanding and provide new guidance on two questions: First, how should we think about and measure biases in AI systems, consistent with the new understanding? Second, what kinds of bias in an AI system should we accept or even amplify, and what kinds should we minimize or eliminate, and why? The key to answering both questions, we argue, is to understand biases as "violations of a symmetry standard" (following Kelly). We distinguish three main types of asymmetry in AI systems – error biases, inequality biases, and process biases – and highlight places in the pipeline of AI development and application where bias of each type is likely to be good, bad, or inevitable.

**Introduction**

The understanding of bias in AI is currently undergoing a revolution. Initially perceived as errors or flaws, biases are increasingly recognized as integral to AI systems and sometimes preferable to less biased alternatives.[1-7] Cognitive psychology and statistics have informed this shift, highlighting the benefits and costs of biases in decision-making processes. Cognitive psychology presents biases as often helpful in making decisions under conditions of uncertainty.[1,5,8-9] Similarly, statistical methods acknowledge biases as often useful and sometimes necessary for making inferences from data.[1-4,6] These insights have been instrumental in redefining biases as not inherently negative, but as sometimes essential components that can and should be harnessed to improve AI systems.

This new perspective on AI biases raises two important questions. First, how should we think about and measure biases in AI systems, consistent with the new understanding? Second, what kinds of bias in an AI system should we accept or even amplify, and which should we criticize, avoid, or eliminate, and why?

**Are Biases Always Bad?**

One might think that whatever biases are, they must be some type of error. One commentator, for instance, defines "[c]ognitive bias" as "errors or flaws in judgment or decision-making, often to the point of denying reality," and "a root cause of medical errors and sentinel events within the healthcare environment."[16] Another notes that "to researchers trained in the psychometric tradition, the … term *bias* is practically synonymous with error, tending to connote general wrongness" (1105).[5]

It's also common to assume that all AI biases are *unfair*. A recent web source, for instance, writes that "Artificial intelligence (AI) bias occurs when a machine learning algorithm makes an error that leads to an unfair result."[30] A recent systematic review defines "bias" as "systematic error in decision-making processes that results in unfair outcomes."[36]

Given these assumed links between "bias," "error," and "unfairness," one might conclude that AI biases are bad and should, as much as possible, be minimized or eliminated: "[b]oth popular and academic articles invariably present algorithmic bias as something bad that should be avoided."[7]

However, as many have recently argued, not all bias is a bad thing, and in fact is often necessary or desirable.[1-6] For instance, a 2022 NIST special report on AI bias concedes that "bias is not always a negative phenomenon"[37]; a recent team of authors write that "bias can be a positive and desirable aspect of a well-engineered model when used to improve other model characteristics"[17] ; and another argues that biases are essential to building AI systems that serve their intended purposes.[1] How can these violations of the usual association between "bias" and "wrongness" be understood?

In our view, the key insight that explains why biases aren't always bad is that the term "bias" is more or less synonymous with "violation of a symmetry standard": wherever anything, including an AI system, exhibits an asymmetry, that asymmetry can be described as a *bias* of a certain kind.[15] Since our world is not composed solely or even primarily of perfectly symmetrical objects and relationships, systems that were completely unbiased would be condemned to inaccurately represent important features of our world, making them less preferable to more biased alternatives that describe the world more accurately.

Further, preferences themselves – including ethical valuations – depend on asymmetries. By failing to encode "asymmetrical" preferences for some things or outcomes over others, unbiased systems are less likely to be effective at guiding human actions in the world and less likely to align with users' values, ambitions, and projects.

However, while a system's exhibiting *some* asymmetries rather than *no* asymmetries is essential to its capacity for success, the precise asymmetries exhibited must be of a sort that improves rather than diminishes the system's accuracy and desirability over alternatives. Otherwise, the system is biased in a *bad* way – that is, its biases carry it off-course from accurately representing the world or effectively guiding our actions within it. These three insights – (1) biases are by definition are just asymmetries, thus not inherently good or bad; (2) in almost all cases, *some* asymmetries in the system are essential to its functioning effectively; and (3) *which* asymmetries a system manifests make all the difference to whether that system performs as desired – are developed further in their application to AI systems below.

**Disciplinary Perspectives on Bias**

Bias is a multifaceted concept that has an important role in many disciplines, including cognitive psychology, ethics, and machine learning. Each of these disciplines offers distinctive insights into how biases function and their implications (Figure 1). For example, cognitive psychology is particularly concerned with "cognitive biases," which it conceives as heuristics or cognitive shortcuts[1,8-11] while ethics emphasizes the moral significance of biases, particularly as indicators of injustice and unfairness.[1,12-13] Statistics generally treats "bias" as synonymous with "a regular pattern of error," but has increasingly recognized the desirability and even functional indispensability of specific kinds of biases, depending on context.[2-4]

The diversity of disciplinary perspectives and theoretical traditions around "bias" is both a strength and a challenge. Hagendorff & Fabi observe that "[o]ver the last decades, 'bias' became a term riddled with ambiguities, especially if one looks into different scientific disciplines and fields."[1] Bias concepts in each field have sometimes enriched the understanding of biases and usefully informed theory and practice in other fields. But the challenge of providing an account of bias that relates and classifies different bias concepts in a clear and coherent way remains largely unanswered – especially if we require of such an answer that it allows for the possibility of *good* as well as bad biases.

| Discipline | Definition of Bias | Key Insights | Applications in AI |
|---|---|---|---|
| Cognitive Psychology | Systematic deviations from "ideally rational" procedures in judgement or decision-making | Biases can lead to errors, but also (in many cases) be cognitive shortcuts that facilitate efficient and accurate decision-making under uncertainty | Understanding cognitive biases can help in designing AI systems that mimic human decision-making processes (for better and worse) |
| Ethics | Deviations from moral standards, often related to fairness and justice | Biases have moral implications, affecting fairness and justice in decision-making | Ensuring AI systems adhere to ethical standards and do not perpetuate unfair biases |
| Statistics | Deviation from a standard, often involving systematic errors in data representation | Statistical biases can be necessary for making inferences from data. Statistical bias alone is not necessarily a sign of error or unfairness. | Identifying biases. Correcting biases where these decrease accuracy or fairness. Providing justification or defense of biases where these increase accuracy or fairness. |

Table 1. Influential Disciplinary Perspectives on Bias

**Bias as Asymmetry: A Transdisciplinary Approach**

Is there anything common to these various approaches to bias? In what sense are they addressing the same or different issues under the same name? Recent discussions by analytic philosophers[1,7,15] point the way to a definition of bias that applies to all or nearly all of these cases, whether "good" or "bad," while retaining the diversity of aspects contributed by a multidisciplinary perspective. This definition is focused on the notion of "asymmetry."[15]

Hagendorff and Fabi propose that "a common denominator for all types of biases is that they can be paraphrased as some kind of distortion, as a tendency towards a particular value, as a specific presetting, or simply as deviation from a standard or a reduction of variety."[1] However, this definition is a compilation of heterogenous features that need not coincide. For instance, a process could involve distortion without tending to a particular value (as in high-noise or "high variance" samples); could tend to a particular value without that value being preset; or could deviate from a standard without reducing variety.

Danks and London give a clearer common formula: "'bias' simply refers to deviation from a standard."[7] This definition's inclusivity is both a strength and a weakness: it can be used to describe cases of all types given in Hagendorff and Fabi's definition, but also includes things that don't fit any common or disciplinary-specific meaning of the word "bias." For instance, any crime would count as a "bias" by this definition since it deviates from the standard of "legal behavior." An athlete who fails to meet a performance standard in their tryout for a team could be described as giving a "biased" performance (even if, say, only 2% of those trying out make the team).

A more adequate definition is suggested by Kelly, who begins by claiming that "bias involves a systematic departure from a genuine norm or standard of correctness" (4).[15] Attributions of bias according to this meaning are always pejorative: that is, they always imply that bias is *bad*. However, Kelly recognizes that our language also includes non-pejorative ascriptions of bias – for instance, the expression "a biased coin." These count as cases of bias, on his view, if they involve the breaking of some contextually relevant symmetry standard.

After recognizing the role of symmetry in non-pejorative bias ascriptions, Kelly comes to recognize the role of symmetry in the pejorative cases as well, suggesting that the extent to which a "genuine norm" violation is classifiable as *bias* correlates to the extent to which the violated norm in question is a symmetry standard: "paradigmatic instances of bias typically involve departures from standards that amount to symmetry violations, while being unbiased involves respecting or preserving certain symmetries and invariances" (153).[15]

For our purposes, we simply adopt Kelly's symmetry-based definition of bias, understanding this as equally applicable to pejorative and non-pejorative uses of the term. Thus, we define bias as "violation of a contextually relevant symmetry standard."

This definition allows us to recognize several important, and too infrequently distinguished, biases (e.g. violations of symmetry standards) in play in artificial intelligence and machine learning contexts. These include asymmetries in how data is collected; asymmetries in the content of that data; asymmetries in how data points are handled by an interpreter (for instance, by an algorithm or theory); asymmetries in how accurately the data represent the phenomena they are taken to represent (whether individuals or populations); asymmetries in how possible interpretations or extrapolations, or theoretical conclusions drawn from the data, are selected (e.g. cases of inability to select one over another due to "underdetermination"); asymmetries between the resultant model and the objects it supposedly represents (e.g. the extent of the model's accuracy); asymmetries in how various parts of the world are represented by the model (i.e. differential accuracy across different parts of the model's representational space); and asymmetries in outputted decisions or classifications (judged by accuracy, demographic parity, or some other metric such as false positive rate or false negative rate across groups).[12]

Two features of this list are especially noteworthy. First, not all of these asymmetries are *always* or even *usually* undesirable: asymmetries in how conclusions are drawn from data, for instance, is unavoidable in conditions of uncertainty[3-4]; and asymmetries in dataset contents (parameters) is desirable where data are supposed to represent an asymmetrical reality. Second, a system can be biased in one of these ways without necessarily being biased in others. These two observations suggest that attributions of bias to a system should generally be accompanied by a specification of (1) what kind of symmetry standard is being violated in this case, and (2) whether the system's violation of that symmetry standard in that context is desirable or undesirable, and why.

Understanding biases as asymmetries newly illuminates common definitions in AI literature by encompassing statistical, cognitive, and ethical perspectives. While statistical and cognitive definitions often focus on systematic deviations from expected values or idealized procedures, with better or worse results for overall accuracy, and ethical definitions tend to emphasize unfair treatment, the asymmetry concept unifies these by highlighting a feature common to all forms of bias: breaking of some possible or expected symmetry.

**AI Biases Are (Usually) One of Three Main Classes of Symmetry Violation**

Understanding biases as symmetry violations allows us to see that the vast majority of what are called "AI biases" belong in one of three main categories of symmetry violation: (1) violations of symmetry between model and representational target (we call these "representational error biases" or simply "error biases"); (2) violations of symmetry in treatment of individuals or datapoints in different groups (call these "unequal treatment biases" or simply "inequality biases"); and (3) violations of symmetry between the processes by which the system operates and an "ideal" set of processes – for instance, deductively valid processes or canons of Bayesian probability (call these "process biases"). (This list might be compared to Hagendorff and Fabi's[1], who recognize three main senses of AI bias in the literature: "cognitive biases," "violation of fairness standards," and "inductive biases." Notably absent from Hagendorff and Fabi's typology are straightforward error biases.)

The traditional association of bias with "a regular pattern of error," and the use of the term "bias" within the field of statistics, mostly match the first of these three categories. When systems are biased in this sense, their outputs or internal components do not represent reality accurately; they deviate from reality in some systematic way. Optical illusions, for instance, constitute biases of these kinds, as do biased samples in statistics. In these cases, the asymmetry is between the represented objects or facts (the "representational targets") and the representing system.

Phrases such as "racial bias" and "gender bias" almost always indicate the second of these categories, "inequality biases." They signal that individuals from some groups (e.g. "African Americans," "women") are treated differently (asymmetrically) by comparison with individuals from other groups (e.g. "Caucasians," "men"). Concerns about "algorithmic fairness" and "implicit bias" are likewise usually concerns about asymmetries of this type. However, *not* all cases of unequal treatment count as unfair or even bad ones; in some cases, we deliberately want the system to treat individuals from different groups differently, because membership in these groups is a relevant differential factor in how the system ought to behave. A financial diagnostic that asks U.S. citizens about their green card status, for instance, is poorly responsive to a relevant difference between individuals. And, in many cases, violation of at least one conceivable "equal treatment" standard is mathematically inevitable, as was demonstrated in the so-called "fairness impossibility" theorem, discussed further below.[31-33]

"Unequal treatment" can also apply to datapoints or theories rather than human individuals. For instance, the argument that learning procedures must be biased to some extent in order to infer conclusions that go beyond their training data, is an argument that at least one "unequal treatment" bias is a necessary condition of any inductive inference.[3-4,22-23]

Finally, the notion of "cognitive biases" best fits the third category: a bias that leads the system to operationally deviate from the performance of an ideal operator. In these cases, the asymmetry is between a processing ideal (say, logical consistency or the canons of probability) and the performance of the system. However, deviations from *one* processing ideal may sometimes involve greater adherence to a *different* processing ideal. In these cases, the system may indeed be "more biased" than competitors according to the first processing standard, but *less* biased than competitors according to the second processing standard. And the extent of bias or lack of bias of this type is not a guarantor of bias in the other two main senses (inaccuracy or unequal treatment). These observations echo the somewhat paradoxical result of decades of research into cognitive "heuristics and biases"[10]: namely, that such biases bring a variety of benefits and costs to cognitive processes, depending on the circumstances in which they're employed.[5,8-9]

Bias in "bias-variance tradeoffs," discussed further below, likewise concerns ways that data processing deviates from some processing ideals. But, like cognitive biases, these deviations sometimes bring the system closer to *other* processing ideals that are ultimately more contributory to accuracy than those that are violated. (And the "bias" in such systems is often itself a source of *error* bias – just not, in cases where more "bias" is recommendable over more "variance" – as *much* a source of error as the variance that is thereby reduced.)

These three main types of bias in AI-ML systems can be summarily visualized by the images in Figure 1, which stress the "broken symmetry standard" that defines each type of bias. The three types of bias can also usefully be conceived as related to one another along two axes of comparison: inaccuracies versus inequalities; and processing biases versus static content (input and result) biases (Figure 2). Finally, common biases of the three types can be identified at various stages in the lifecycle of AI-ML development and deployment (Figure 3).

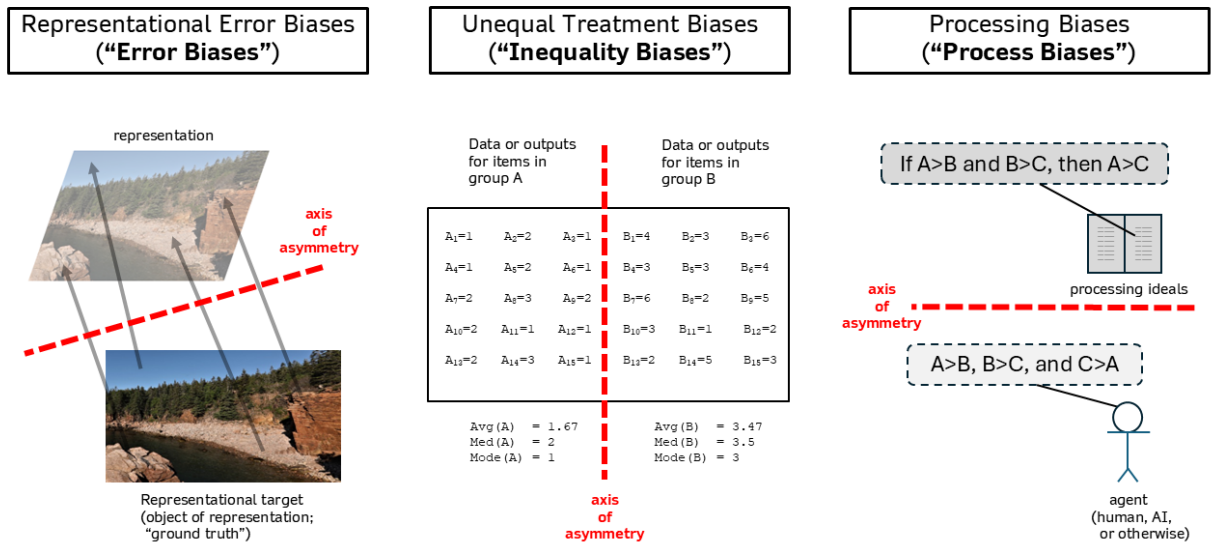

Figure 1. Biases in AI-ML Systems as Symmetry Violations: Three Main Types.
Photo credit: Sardius Stalker, Acadia National Park, National Park Service.

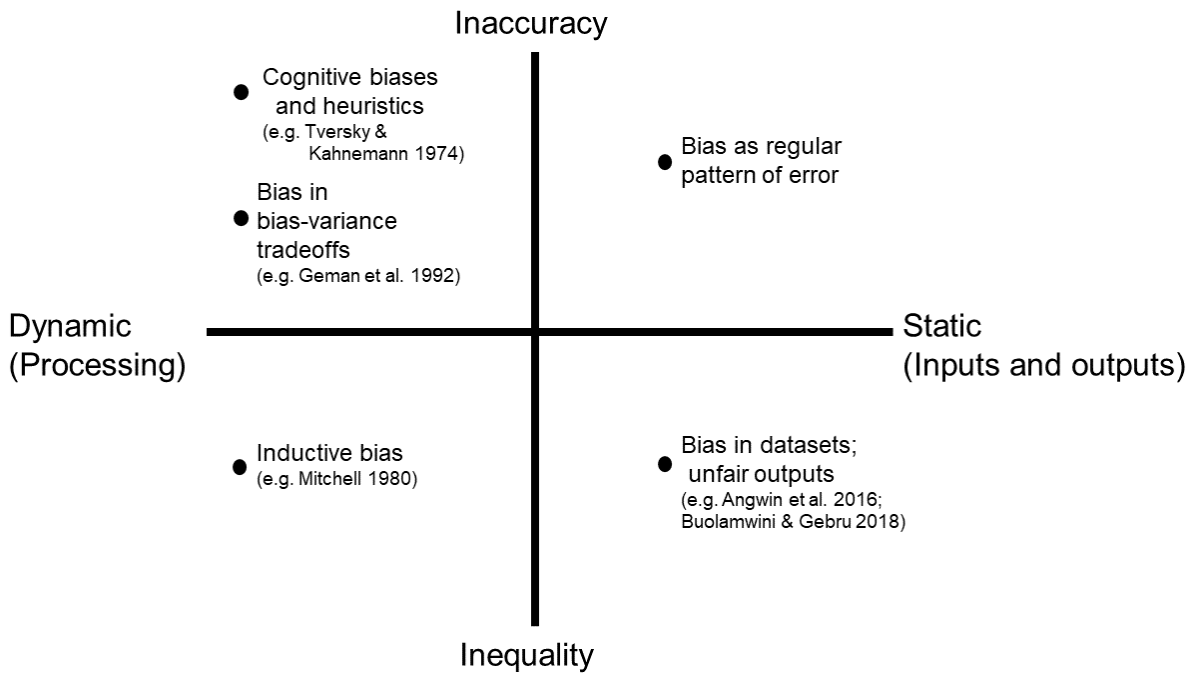

Figure 2. Common AI-relevant biases, classified along two dimensions.

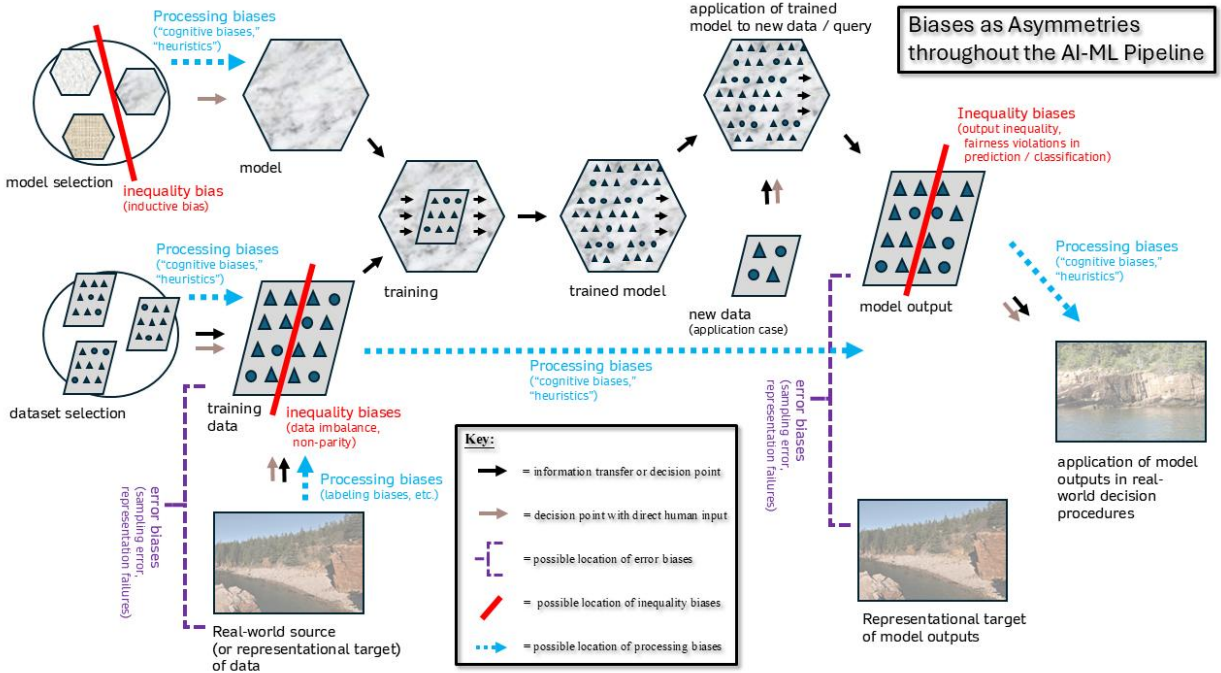

Figure 3. Biases as Asymmetries throughout the AI-ML Pipeline. Photo credit: Sardius Stalker, Acadia National Park, National Park Service.

**The Ambiguous Goodness of Common AI Biases**

In this section we look more closely at a few commonly discussed types of AI bias. We interpret these in terms of the type of asymmetry they exhibit, and show that their "goodness" or "badness" can be better understood from this perspective.

*(a) Cognitive Biases*

The literature on "cognitive biases" describes processes of reasoning or inference that have suboptimal results in many cases, or deviate from canons of rationality – for instance: principles of first-order logic or Bayesian probability theory. To understand what kind of bias is highlighted by the term "cognitive bias" and the closely related (sometimes synonymous) "heuristics," we must review a bit of the history of these terms.

Since Tversky and Kahneman, the terms "cognitive bias" and "heuristics" have often been used interchangeably.[10] "Cognitive bias" carries some implication of error or at least deviation from optimal rationality, whereas "heuristics," in its use by Simon and others, implies a shortcut that is nonetheless effective at problem solving.[18] Recent research has shown that certain cognitive biases can enhance AI performance in specific contexts.[5] For example, the "availability heuristic" bias, which prioritizes easily recalled information, can be beneficial in AI systems designed for rapid decision-making in emergency situations.[38]

One might suppose heuristics can contribute efficiency but only at the expense of accuracy. However, some have argued that cognitive biases and heuristics are actually optimal for *accuracy* in some circumstances, particularly those in which the application context is very different from the training context.[8-9,11, 28] The relative benefits and costs of a more biased system, in these circumstances, are similar to those between higher-bias-lower-variance systems and higher-variance-lower-bias systems in bias-variance tradeoffs (discussed further below).

It is important to note that, in cases where higher bias is found to be beneficial, not just any bias will achieve the benefit. The bias must be in a direction and of a type and quantity that improves overall accuracy (even if the system is still inaccurate in many of its particular decisions or predictions, and even if it is this way *because of* the bias). This means that *some* cognitive biases, in *some* circumstances, reduce "error bias," and are therefore "good" or "preferable" in those circumstances. But not *every* cognitive bias is good or preferable in *every* circumstance. The particular cognitive bias *and* the particular context are of essential importance in determining whether a particular cognitive bias is an overall beneficial feature of a system.

(b) *Discriminatory biases; unfair biases; unjust biases*

Concerns that a decision-procedure is biased for or against some decisions for individuals in some groups in comparison with individuals in others, are concerns about bias as unequal treatment. Terms like "racial bias" and "gender bias" fit this description, as do most discussions of algorithmic fairness and fair AI-ML. Well-known statistical tests for fairness such as demographic parity, false positive rate parity, and predictive accuracy parity,[12-14] are often used to estimate the extent to which biases of these kinds characterize a system's decision-making. These metrics differ from one another in what *kind* of equality or inequality of treatment they measure (for instance: equal distribution of a trait between groups [*"independence"*]; equal rates of false positives or false negatives between groups [*"separation"*]; or equal predictive power of assigned scores to individuals in both groups [*"sufficiency"*]).[12-14]

A paradoxical feature of these fairness metrics is that, except in a rare and narrow set of circumstances, they cannot be simultaneously satisfied.[28] Thus, in most cases it is necessary for a decision-making system to be unfair by at least one standard of fairness that legitimately applies to some cases. Since no-one can be asked or expected to do the impossible, the fairness impossibility result highlights the need to be careful and deliberate about choosing the fairness standards by which one evaluates a system.[29] AI developers should carefully consider which fairness criteria are most appropriate for their specific use case and be transparent about the tradeoffs involved. For instance: proportional parity ("independence") is likely preferable in circumstances where resources should be allocated completely independently of subject features, such as distribution of stimulus payments to individual citizens in a stressed national economy. Reduction of false positive or false negative rate imbalance ("separation") may be preferable when false positives or false negatives are particularly damaging or ethically unacceptable, such as deciding on anti-suicide interventions or eligibility for pre-trial release or a loan. Maximation of predictive accuracy parity ("sufficiency") may be the best fit when overall accuracy is the primary target, such as advertising or recommender systems.[29]

It's extremely important to recognize that not *all* cases of *unequal* treatment are cases of *unfair* treatment. For instance: "Biased" allocation of medical resources based on differences in risk factors among patients violates the fairness standard of proportional parity ("independence"). Yet this "biased" allocation may be exactly what we want the system to do, since we want it to allocate resources to those who need them and not waste resources on those who don't. In these cases, minimizing "error bias" and achieving our goals for the system can *only* be accomplished through "unequal treatment." Unequal treatment is not by itself equivalent to unfairness.

(c) *Inductive biases*

Mitchell[3] provides an early example of an argument that biases are necessary for informative inductive inferences (though his reasoning parallels that of Goodman[23] and Hume[22]). Mitchell writes that "we use the term bias to refer to any basis for choosing one generalization over another, other than strict consistency with the observed training instances."[3] He then argues that consistency with the observed training instances will never be enough to select among alternative generalizations within what he calls the "version space" of possible generalizations. The generalization that is identical to the training instances, on the other hand, provides no means to predict any data points beyond what's already given in those instances. Thus, biases are recommendable, even necessary, as means to break the tie between otherwise equivalently possible generalizations within the version space. Mitchell points to some specific kinds of biases that can be useful for selecting a generalization, such as "[f]actual knowledge of the domain," "[i]ntended use of the learned generalizations," "[k]nowledge about the source of training data," "[b]ias toward simplicity and generality," and "[a]nalogy with previously learned generalizations."[3] Mitchell's discussion focuses on systems that "learn" to generalize on the basis of training data, but a parallel argument can be made for any inductive system.

Subsequent arguments for the necessity of bias in inductive inferences follow a similar trajectory as Mitchell's, often with citations to his 1980 paper.[3-4,24]

Notice the very limited sense of "bias" according to which such systems are shown to be biased in these arguments: namely, they are biased in their selection of an inductive procedure or conclusion *from among all possible procedures or conclusions consistent with the data alone* (the "model selection" step in Figure 3). If, on the other hand, we examine the choice of inductive procedure from a standpoint that *includes* the supplementary information (for instance, domain knowledge), then the choice may be very *un*biased by relevant *processing* standards, insofar as it tracks relevant differences between the available hypotheses.

Likewise, if we consider these "biased" inductions by the standard of accuracy to their representational targets – that is, their "error bias" (in Figure 3, this is the error biases associated with "model outputs") – then we see that the "theory-choice-biased" inductions are *more* capable of being representationally *un*biased inductions – a fact implied by the authors' description of theory-choice-biased inductions as potentially *more accurate* (that is, more accurate as descriptors or predictors of their representational targets).

Furthermore, while these arguments show that biased inductions *can be* more accurate than completely unbiased ones (insofar as completely unbiased inductions fail to classify or predict future instances that differ from the training instances in any way), they don't show that more biased inductions are *in general* more accurate than less biased ones; nor even that *all* biased inductions are an improvement over completely unbiased ones. Consider that an induction that misclassifies or falsely predicts future datapoints is hardly better, and possibly worse, than one that refuses to classify or predict at all. At least one possible generalization in what Mitchell calls the "version space" will have this "gets *every* new datapoint wrong" feature; thus, at least one available "biased" induction is arguably as bad or worse than a completely unbiased induction.

(d) *Bias in bias-variance tradeoffs*

The "bias-variance" tradeoff has long been known to machine learning practitioners.[1-2,6] Here "bias" means "systematic deviation of predicted values from actual values" and is contrasted with "variance," which signifies a random deviation of predicted values from actual values. The discovered "trade-off" is that sometimes increasing the bias can decrease the variance, and sometimes this can result in lower overall inaccuracy than if the bias were lower but variance were higher. In other words: Sometimes higher-bias models exhibit lower variance than lower-bias alternatives, and this reduction in variance can lead to higher overall accuracy for the model.[2,6,8]

Geman et al. introduced the idea of a bias-variance trade-off.[2] Beginning with a contrast between *non-parametric* ("model free") and *parametric* ("model dependent") approaches to statistical inference tasks, they argued that the estimation error of a statistical inference procedure could be decomposed into two main components: bias and variance. Roughly put, variance is random and non-directional error, whereas bias is systematic error (typically directional, though perhaps in different directions for different parts of a dataset or learning task). Model-free architectures tend to have low bias but high variance, while model-dependent architectures exhibit lower variance but higher bias. Since all statistical inference procedures are somewhere on a spectrum between model-free and model-dependent, all such procedures can be located in a space of tradeoffs between bias and variance.

In some circumstances, the risk and extent of estimation errors arising from variance is greater than that arising from bias. Introducing bias into such systems can reduce variance and, in many cases, reduce overall error as a result. Simple examples of such variance-reducing but bias-increasing procedures include eliminating outliers; using drop-out[27]; reducing the parameter size of the resulting model (e.g. requiring the model function to be no greater than an n-term polynomial); and stopping the training procedure before overfitting occurs. In circumstances where more error derives from variance than bias, such procedures can improve predictive accuracy overall.

Features of learning environments wherein variance is a major challenge to accuracy include: low quantity of training data; noisy data; and prediction or application contexts that are relatively unlikely to be like the training context.[8-9] In these environments, a system that overfits will be drastically out of step with its representational target for any new applications (interpolations, extrapolations, or otherwise). A more biased system, however, can keep this variance problem in check by holding predictive values within a range less likely to be wildly inaccurate.

This line of thinking supports the conclusion that more biased inductions are *sometimes* more accurate than less biased ones. However, several qualifications of this conclusion should be noted.

First, *the precise bias that is selected* makes an enormous difference. In a case where variance is a problem and introduction of biases can help, some biases that could be selected will reduce variance more than they generate new sources of error. But, in almost all such cases, some others of the biases that could be selected will reduce variance but *also* generate greater sources of error than they eliminated (now due to bias). Without any reason to choose one bias over another, we risk making things worse rather than better with any bias we choose (for the same reasons noted earlier in connection with "cognitive biases" and "inductive biases"). Further, when we have a reason to choose one bias rather than another, this *choice* of bias is itself evidentially motivated, and hence *not* "process biased" in a problematic way.

Second, *the benefit of the bias in bias-variance tradeoffs is merely relative and instrumental*: if a model that were equivalently lower variance but *also* lower bias became available, that lower-bias-and-lower-variance model would generally be preferable to the high-bias model. This shows that the epistemic goodness of increases in such biases is merely an instrumental goodness (tied to reduction in variance).

**Can Biases be of More Than One Type?**

Biases sometimes fit into two or more of the main types simultaneously. For instance, suppose a non-profit organization providing food assistance has explicitly adopted, as a processing standard, that all incoming applications for emergency assistance should be ranked first by severity of need, then by proximity to the non-profit's storage facilities, in determining order of response. A review of the organization's practices reveals that all applications from some locations were immediately placed at the bottom of the list, regardless of severity of need. This is a violation both of the stated processing ideal and of the principle of equal response for all applications. (The processing ideal in this case *is* a principle of equal response; hence the violation is both a case of inequality bias and a case of processing bias.)

To take another example: suppose a model of consumer preferences is trained on a very imbalanced dataset, such that consumers in some demographic groups (e.g. women) are represented in a way that deviates significantly from the actual distribution of behavior in the larger population of that demographic group. The deviation itself is a case of sampling error, and thus a case of representational error. But the difference in accuracy between men and women that results constitutes a case of inequality bias as well.

Finally, consider a case that combines processing biases and representational error biases. Suppose John is the concierge at a hotel and notices that many of the people checking into the hotel have medical licenses. He reasons that because all practicing doctors have medical licenses, these individuals must be practicing doctors. Suppose further than most of those checking into the hotel are not practicing doctors, but rather pharmaceutical researchers attending a pharmaceutical research conference, who need the licenses to do research. In this case John violates a processing ideal of rational inference ("affirming the consequent" fallacy: "If A then B; B; therefore A"), thereby exhibiting a processing bias. But he also ends up with a belief that constitutes a representational error bias regarding the professions of the individuals at the conference.

If cases of bias can fall into more than one of these three categories, does the distinction between them still hold value? We believe that it does. Distinguishing and identifying the various types of asymmetry involved in a particular case of bias is essential to determining (a) which of these asymmetries are ethically unacceptable, ethically acceptable, or even ethically desirable or required; and (b) identifying which features of these cases ought to have been handled differently or should be mitigated posthoc in order to address the biases that are determined to be problematic. To take an analogous case: There are different kinds of "rights" that humans can have (for instance: legal rights and human rights). Many cases that involve a violation of one of these rights also involve violations of others. Nonetheless, it is analytically valuable – and in some cases essential – to distinguish different kinds of rights that have been violated in a particular case, as a step towards identifying the actions that should be taken to protect and restore those rights.

## When Are AI Biases Good or Bad, Necessary or Unnecessary?

Previous discussions of AI bias often classify biases by the stage of the AI lifecycle at which they appear: for instance, biases in datasets, in model selection, in output, and in applications.[7] Combining this common approach with the asymmetry-based classification of AI biases developed in previous sections provides a novel systematic framework for understanding and evaluating AI biases (Figure 3). This framework enables us to more clearly distinguish biases that are *necessary* from those that are *unnecessary*, and biases that are *desirable* from those that are *undesirable*, as well as to think about trade-offs between biases of different types (Table 2).

| HORIZONTAL: Stage -------------- (System Type) -------------------------- VERTICAL: Type of Symmetry Violation | Model selection ----------------- (Human) ----------------- | Inputs (e.g. datasets) -------------------- (Human, AI, Human + AI, Other) -------------------- | In-processing (e.g. training, predicting) ------------------- (AI, Human + AI) ------------------- | Outputs (e.g. outputted predictions) ----------------- (AI) ----------------- | Applications ----------------- (Human + AI) ------------------- |
|---|---|---|---|---|---|
| (1) **Error biases** | X | Bad (mostly), but see (d) below | X | Bad (mostly), but see (d) below | X [Except in cases of "knowledge discovery"] |
| (2) (b1) **Discriminatory biases** ("proportional distributive parity") | X | Good or bad  Necessary (usually) | Good or bad  Necessary (usually) | Good or bad | Good or bad |
| (2) (b2) **Discriminatory biases** ("false positive rate balance," "false negative rate balance," and "accuracy parity") | X | X | X | Bad (usually)  Necessary (in some regard, usually) | Bad (usually)  Necessary (in some regard, usually) |
| (3) (a) **Cognitive biases and heuristics** | Bad (mostly), but see (d) below | X | Bad (mostly), but see (d) below | X | Bad (mostly), but see (d) below |
| (2) and (3): (c) **Inductive biases** | Necessary (in some regard) | X | Necessary (in some regard) | Necessary (in some regard) | Good or bad |
| (1) and (3): (d) **Bias in bias-variance trade-offs** | Bad (mostly), Good (sometimes) | X | Bad (mostly), Good (sometimes) | Bad (mostly), Good (sometimes) | Bad (mostly), Good (sometimes) |

Table 2. A classification of AI biases by stage of AI development cycle in which they appear (*horizontal axis*), and type of asymmetry they exhibit (*vertical axis*). An "X" indicates that that type of asymmetry isn't intuitively ascribable to what happens at that developmental stage. (1) = error biases; (2) = inequality biases; (3) = process biases.

|  | Examples of (Potentially) Good Biases | Examples of (Potentially) Bad Biases |
|---|---|---|
| Model Selection | Selecting one model over others based on expectations of higher accuracy or fit-to-performance-goals of the model (i.e. inductive bias).[3] | Selecting one model over others based solely on ease of use or prior familiarity with the model. |
| Data Production | Adoption of clear and consistent guidelines for data-labeling that produce datasets well-suited for a specific analytic goal. | Use of data collection, labeling, or preprocessing practices that result in (a) under-representation of some groups (unfair inequality biases) or (b) systematic deviation from accuracy across the dataset or across data of specific types. |
| Dataset Selection | Intentional oversampling of under-represented groups to create a more balanced dataset that reduces error biases and/or harmful inequality biases. (The good bias in this case is probably best classified as an *inequality bias* and a *processing bias* insofar as it involves treating datapoints from different groups differently, a procedure that would usually not be ideal but in this case is adopted for good reason.) | Selection of datasets that are highly imbalanced between demographic groups, leading to error biases and/or inequality biases in outputs . |
| In-Processing | Adjusting an algorithm to give slightly more weight to features that have been historically undervalued but are known to be important. | Use of an opaque deep learning model that inadvertently amplifies societal stereotypes. |
| Model Outputs | Outputs are asymmetrical across various categories, but this asymmetry is (a) accurate to the reality they are intended to track and (b) doesn't include inequality biases that go on to play a role in decisions causing inequitably distributed harms | Outputs are (a) inaccurate to the reality they are intended to track, and/or (b) include inequality biases that go on to play a role in decisions causing inequitably distributed harms. |
| Applications | Decisions made are informed by the AI-ML system's outputs in a way that shows appropriately different responses in relevantly different situations and application contexts | Decisions made are informed by the AI-ML system in a way that demonstrably subjects some groups or individuals to greater harms than others, without any good justifying reason for the differences. |

Table 3. Examples of good and bad biases at each stage of the AI development and application cycle

Figure 3 and Table 2 gives a classification of AI biases by (a) the type of asymmetry that defines them and (b) the stage of the development cycle at which they appear. This allows us to note which types of biases are unavoidable (that is, necessary), as well as which are (usually) good or (usually) bad. Some stages of the development cycle provide little or no opportunity for biases of particular kinds; in these cases, we've indicated the absence with an "X."

We can separate biases defined by these two axes into three categories: *Necessary (in some regard)*; *Usually good*; and *Usually bad*. It is important to distinguish between claiming that *some* bias of the specified type is "necessary," "usually good," or "usually bad"; and claiming that *a specific bias* (that is, a bias *in a specific direction* and *to a specific extent*) is "necessary," "usually good," or "usually bad."

(A) *Necessary biases*

*Inductive biases* in the stages of model selection, in-processing, outputs, and applications are almost always a necessary condition of carrying out an induction (a prediction or classification) at all. For the reasons discussed above, such biases are necessary for inductive inferences of any kind. However, the *precise* biases selected or exhibited make a big difference to whether the results will be good or bad. In this sense, some specific inductive biases are better than others, and some are quite bad. Which specific biases are better or worse will usually depend heavily on features of the context.

Likewise, *unequal treatment* in datasets, in-processing, outputs, and applications is often necessary as well, for two basic reasons. First, datasets often contain evidence of bias in the systems they provide information about. Indeed, a completely symmetrical dataset is likely to be relatively uninformative. One way to see this is that the more symmetries characterize a dataset, the more redundant is the data in that dataset: the usual gains from larger datasets in comparison with smaller would, in many cases, not be garnered here, since the larger dataset will have the same structural features as the smaller one. Second, when it comes to in-processing, outputs, and applications, the fact that "fairness" can be measured by competing and incompatible metrics ("fairness impossibility") means that at least one of the incompatible metrics will be violated in almost all cases.

What should be done about unavoidable biases of these two types?

For inductive biases, we recommend (a) accepting that some inductive bias is unavoidable, but (b) trying to select the inductive bias that maximizes accuracy, fairness, and other goals of the induction.

For biased datasets, we recommend evaluation of the biases in light of the functions the datasets will be used for. For instance: A model designed to accurately measure an asymmetrical reality generally *ought* to be biased in the same way and to the same extent that the represented reality is "biased." However, when models are used to inform decision-making at the "application" step, such biases can provide grounds for differential treatment that is ultimately unfair. Bias in datasets must therefore be handled delicately, in context of the full use cycle of data selection, training, and deployment.

In general, unequal treatment biases at the in-processing, outputs, or applications stages should be vetted via the fairness metrics that appropriately apply to them. This requires carefully selecting the metrics that should apply in each case, and, in cases where these are mutually incompatible, carefully deciding which to prioritize and to what extent.[28-29]

(B) *Usually good biases*

In general, biases of an AI are "good" (when they are) for one of three reasons:

(1) the bias in the AI system *represents* a bias in the system modeled (i.e. the system's bias effectively tracks asymmetries in the world that the AI system is being used with the intention of tracking). Examples include inductive biases, cognitive biases, and some inequality biases.

In these cases, the bias in that part of the model should be preserved; however, the added value of the bias for *representation* purposes should be quarantined from any unfair effects on *actions* (recommended or selected).

(2) the bias in the AI system *reduces noise* in the modeling of the system modeled (i.e. reduces *variance*), thereby improving accuracy, as recognized in the bias-variance trade-off. In these cases, the selection of a higher bias model may be justified; but the researcher should also (i) continue to explore the possibility of simultaneously lower-bias and lower-variance models; and (ii) seek an explanation for the success of the higher-bias model in this case that can help to guide the search for simultaneously lower-bias and lower-variance models.

(3) the bias in the AI system tracks *preference* asymmetries – for instance, by ranking items of some kinds higher than others, or by differentially responding to individuals in a manner than corrects for or repairs previous unfairness biases. These can be described as process biases and (often) as inequality biases, but ones that track adopted preferences.

In all of these cases, we encourage preservation or amplification of the useful bias, so long as attention to other possible biases resulting from the current model and their total costs (as well as the costs of other possible undesirable effects of the model) are borne in mind. The useful bias should not create more serious problems than it solves.

(C) *Usually bad biases*

In general, biases of an AI are "bad" (when they are) for one of three basic reasons:

(1) the bias in the AI system *distorts or deviates from an accurate representation* of the systems it is used with the intention to represent;
(2) the bias in the AI system *distorts or deviates from a fair representation* of the systems it is used with intention to represent;

(3) the bias in the AI system comes at a *cost to other desirable features* of the AI system (e.g. its efficiency; the trust assigned to it by the public; etc.)

In cases of "bad" biases by any three of these standards, the developer has a number of options, including:
      (i) *eliminate* the bias (via supplemented *datasets* or *in-processing* procedures)
      (ii) *mitigate* the bias (via post-processing at end of in-processing or the outputs stage; or an adjusted application procedure)
      (iii) *accept* the bias as a "best case scenario" or "necessary evil" (for instance: if an analysis in terms of "bias-variance" tradeoff reveals it to be so).

**Outstanding Questions and Opportunities**

Future research directions for a symmetry-based approach to AI biases include (i) development of quantitative measures for each type of bias identified in the typology; (ii) investigation of relations of amplification or trade-off between different types of biases; and (iii) exploration of how this typology can inform regulatory frameworks for AI governance. We leave these tasks for future efforts.

Some newly discussed types of AI bias haven't been covered in this review. These include *temporal biases* (AI systems trained on historical data may struggle to adapt to rapidly changing environments or societal norms); *multimodal biases* (as AI systems increasingly integrate multiple types of data such as text, images, and audio, new forms of bias may emerge from the interactions between these modalities); and *feedback loop biases* (AI systems deployed in real-world settings may create self-reinforcing biases through their interactions with users and environments).

In addition, when models are fine-tuned for new tasks, biases from the original task can transfer, leading to a variety of what may be called *transfer learning biases* such as cultural bias propagation (e.g. when a model trained on English text is fine-tuned for other languages, it may transfer Western cultural biases), domain-attachment biases (e.g. a model originally trained for medical diagnosis might transfer biases when fine-tuned for a different medical task), or other unexpected associations (e.g. example, a model trained on news articles might associate certain topics with positive or negative sentiment based on how they were portrayed in the news rather than their inherent emotional content).

There are also a variety of *federated learning biases* that can emerge from the use of local datasets that may be relatively heterogeneous in comparison with one another. For instance, local datasets at different nodes may have different distributions or characteristics that could lead to Non-IID (Independent and Identically Distributed) data. If not all nodes participate equally in the training process due to factors like connectivity issues or resource constraints, this can lead to underrepresentation of information at some nodes in comparison with others. In federated learning between local models, divergence between local model structures, or between local and global models, can generate biases as well.

While we pass over the full task of analyzing these new forms of bias within our framework, Table 4 gives a brief suggestion of how each emerging type of bias might be classified and understood.

| Type of Bias | Type of Asymmetry | Axis of Asymmetry |
|---|---|---|
| Temporal Biases | Inequality | Between past data and present or new data |
| Multimodal Biases | Inequality | Between data, in-processing of data, or outputs in one modality, and data, in-processing of data, or outputs in another |
| Feedback Loop Biases | Process | Between an ideal of accuracy or process neutrality, and the prior weights, information, or decisions of the system |
| Transfer Learning Biases | Inequality, error, or process | Between an ideal of processing, equal treatment, or accuracy, and the behavior of the new (transferred-to) system |
| Federated Learning Biases | Inequality, error, or process | Between an ideal of processing, equal treatment, or accuracy, and the local or global behavior of the system |

Table 4. Emerging types of AI biases understood as asymmetries

**Is Inequality Bias the Most Basic Kind of Bias?**

There are reasons to think that inequality bias is an essential component of all biases -- that is, a kind of asymmetry that *must* be present in order for a system to count as "biased" in any way.

First, consider cases of representation bias without inequality bias. Such cases might involve random (i.e. symmetrically distributed) error – but, in that case, the error is usually *distinguished* from bias and called "variance" or "sampling error" instead. Or such cases might involve a regular and perfectly symmetrically applied transposition of values from the represented object or system to the representing object or system; but a perfectly symmetrically applied "transposition" appears to be a feature of *all* representations, even entirely or especially accurate ones. Retrieving information loss from such a transposition is usually just a matter of reversing the transposition.

If, however, we introduce *different standards of treatment* of different parts of the represented object or system – that is, an *inequality bias* regarding how the object or system is represented – then the ascription of "bias" to the representing system comes much more naturally. Notice that transformations that involve (say) adding a single value to all data points would count as an inequality bias in at least one sense, since the addition adds a different proportional value to data points of different sizes (i.e. it is biased *in favor* of raising the proportional value of smaller data points and *against* raising the proportional value of larger data points).

Second, consider cases of process bias without inequality bias. These cases involve deviation of a process from a processing idea, but if the ideal itself *isn't* a symmetry standard – such as, "give all possible answers equal consideration" (violated by the availability heuristic) – then the description of this failure of the processing standard as a "bias" seems incorrect.

If inequality bias is indeed the fundamental "type" of bias, and all others are an instance of it, why might this fact go unnoticed for so long? One reason is the tendency to assume that "unequal treatment" only describes circumstances in which people are treated unequally; another is the tendency to assume that unequal treatment (whether of people or other things) is always bad. When we insist on thinking and speaking more clearly about these issues, the central and defining role of "inequality biases" in all types of bias may be easier to see.

Despite the centrality of inequality biases, we believe the threefold typology offered here still has value. The relevant distinction is between biases that distort a *representation* away from a more accurate description of reality; biases that involve assigning *resources* in different amounts or proportions between possible recipients (whether people, as in the fairness literature; or theories, as in the literature on inductive bias); and biases that characterize a *decision-making process* and show up as a failure to respond in the same way to cases that are alike in some salient way.

Biases, then, appear in *representations*, in *resource distributions*, and in *treatment of data* within cognitive or computational processes, because these are three distinct but important types of "equal treatment" standards that humans apply to themselves and to AI.

**Conclusions**

The enhanced typology of AI biases proposed in this paper has several implications for AI development and evaluation. By categorizing biases based on both their lifecycle stage *and* asymmetry type, developers can better identify and address biases at each stage of AI development, as well as distinguish necessary from unnecessary, and good from bad biases. This approach not only aids in identifying and mitigating harmful biases, but also in optimizing positive biases to enhance AI performance and fairness. For instance, distinguishing representative from anti-representative biases in datasets can guide the development of more representative datasets. Distinguishing the ineliminable elements of inductive biases and unfairness biases from the optional and variable features of these reduces the risk of accepting harmful biases that might otherwise have been avoided. Finally, acknowledging tradeoffs between process and inequality biases of different types clarifies that choices about processes and about which symmetries to preserve or disrupt are almost never selections between "biased" and "unbiased" processes, but rather between processes that are biased in different ways, some of which are beneficial and some of which are harmful, in a variety of ways and dependent on the context.

**Acknowledgements**

The authors thank the Center for Equitable AI and Machine Learning Systems (CEAMLS) for their support of this research.